\documentclass[aps,onecolumn,showpacs,showkeys,nofootinbib]{revtex4}
\usepackage{epsfig}
\usepackage{amsmath}
\usepackage{amsfonts}
\usepackage{amssymb}
\usepackage{graphics}
\usepackage{colordvi}

\begin{document}

\title{Galaxy rotation curves in $f(R,\phi)$-gravity}

\author{A. Stabile$^1$\footnote{arturo.stabile@gmail.com}, S. Capozzielo$^{2,3}$\footnote{capozziello@na.infn.it}}

\affiliation{$^1$Dipartimento di Ingegneria,
Universita' del Sannio, Palazzo Dell'Aquila Bosco Lucarelli, Corso Garibaldi, 107 - 82100, Benevento, Italy,}
\affiliation{$^2$Dipartimento di Scienze Fisiche, Universit\`{a} di Napoli "Federico II",
 Compl. Univ. di Monte S. Angelo, Edificio G, Via Cinthia, I-80126, Napoli, Italy}
\affiliation{$^3$ INFN Sez. di Napoli, Compl. Univ. di Monte S. Angelo, Edificio G, Via Cinthia, I-80126, Napoli, Italy.}

\begin{abstract}

We investigate the possibility  to explain theoretically  the galaxy rotation curves by  a gravitational potential in total absence of dark matter. To this aim an analytic  fourth-order theory of gravity,   nonminimally coupled with a massive scalar field is considered. Specifically, the interaction term is given by an analytic  function $f(R,\phi)$ where $R$ is the Ricci scalar and $\phi$ is the scalar field.  The gravitational potential is generated by a point-like source and compared with the so called Sanders's potential that can be exactly reproduced in this case. This result means that  the problem of dark matter in spiral galaxies could be fully addressed by revising general relativity at galactic scales and requiring  further gravitational degrees of freedom instead of new material components that have not been  found out up to now. 
\end{abstract}
\date{\today}
\maketitle

\section{Introduction}

J.H. Oort first pointed out  the {\it missing matter problem} in the 30's  of last century \cite{Oort1,Oort2}. The issue came out 
by observing the Doppler shift  of stars moving near the plane of our Galaxy and calculating   the star velocities. The result was  that there had to be a large amount of  matter inside the galaxy to prevent  the stars from escaping. Such a "matter" should give rise to a central gravitational force  much larger than   Sun's gravitational pull to keep  a planet in its orbit.  However  it turned out that there was not enough luminous mass in the Galaxy to account for this dynamics.  The discrepancy was very  large  and  the Galaxy had to be at least two or three times more massive  than the  sum of  all its luminous  components in order to match the result.
Later on,  the tangential velocity of stars in  orbits around the Galactic center was calculated as a function of distance from the center.  Surprisingly it was found that far away   from the Galactic Center,  stars move with the same velocity independent of their distance out from the Galactic Center. These results strongly posed the problem that either luminous matter was not able to reliably trace the radial profile of the Galaxy or the Newtonian potential was not able to describe dynamics far from the Galactic center.

Soon after,   other dark matter  issues  came out  from   dynamical descriptions of  self-gravitating  astrophysical systems like stellar clusters,   galaxies, groups and  clusters of galaxies.
 In all these  cases, there is more matter  dynamically inferred than that can be accounted for by luminous matter components. The mass discrepancy comes out  assuming the validity of  Newton law  at any  astrophysical scales. Problems emerged also  at  larger scales.  F. Zwicky discovered anomalous motions of galaxies in  the Coma cluster finding that  the visible mass  was  too little to produce enough gravitational force to hold the cluster together  \cite{Zwicky}.

At the beginning, the only possibility considered was to assume the  Newton law holding at all scales  and  postulating  some non-luminous component to make up the missing mass.    Many names have been coined to define these invisible components.
For example, the MAssive Compact Halo Objects (MACHOs) are objects like black holes  and neutron stars (in general sub-luminous objects) that  populate the outer reaches of galaxies like the Milky Way.  There are the Weakly Interacting Massive Particles (WIMPs) which do not interact with standard matter  (constituted by baryons as protons and neutrons):  they are supposed to be particles out of the Standard Model of Particles but, up to now, there is no final indication for their existence  \cite{revdonne}.  In general, dark matter is assumed  to come in two flavors, hot (HDM) and cold (CDM) dark matter.
 The CDM should  be in dead stars, planets, brown dwarfs etc., while HDM should be constituted by   fast moving relativistic particles. It should be  neutrinos, tachyons etc. However,  there is still no definitive proof that WIMPs exist, or that MACHOs will ever make up more than five percent of the total amount of missing matter.

On the other hand, the need of  unknown components as dark energy (coming from cosmology)
and dark matter  could be considered nothing else but as a signal
of the breakdown of Einstein General Relativity (GR) at astrophysical
(galactic and extragalactic) and cosmological scales.

In this context, Extended Theories of Gravity (ETGs) could be, in
principle, an interesting alternative to explain cosmic
acceleration  and large scale structure
without any dark components. In their simplest version, the Ricci
curvature scalar $R$, linear in the Hilbert-Einstein action, could
be replaced by a generic function $f(R)$ whose true form could be
"reconstructed" by the data. In fact, there is no a priori reason
to consider the gravitational Lagrangian linear in the Ricci
scalar while observations and experiments could contribute to
define and constrain the "true" theory of gravity (see \cite{PRnostro,reviewodi,reviewodi1,reviewmauro,reviewvalerio,libro,libro1}).

Coming to the weak-field limit,  any alternative relativistic theory of gravity is expected to
reproduce GR results which, in any case, are firmly tested only at 
Solar System  scales in the Newtonian limit \cite{Will93}. Even this limit  is 
matter of debate since several relativistic theories  do not 
reproduce it. For example, 
Yukawa-like corrections to the Newtonian potential easily comes out  \cite{Stelle:1976gc}  with 
interesting physical consequences. For example, it is 
claimed by some authors that  the flat rotation 
curves of 
galaxies can be explained by such
terms \cite{Sanders90}. Other authors  have shown that a
conformal theory of gravity is nothing else but a fourth order
theory containing such terms in the Newtonian limit. 
In general, any relativistic theory of gravitation  yields
corrections to the weak-field gravitational potentials ({\em 
e.g.},  \cite{Qua91}) which, at the  post-Newtonian  level and 
in the  Parametrized Post-Newtonian formalism,
could constitute a test of these theories \cite{Will93}.

This point deserves a deep discussion. Beside the fundamental physics motivations coming from Quantum Gravity and unification theories (see 
\cite{PRnostro,libro}), ETGs pose the problem that there are further gravitational degrees of freedom (related to higher order terms, non-minimal couplings and scalar fields in the field equations) and gravitational interaction is {\it not} invariant at any scale. This means that, besides the Schwarzschild radius, other characteristic gravitational scales could come out from dynamics. Such scales, in the weak field approximation, should be responsible of characteristic lengths of astrophysical structures that should 
result {\it confined} in this way \cite{annalen}. 

In this  paper, without claiming for completeness, we will try to address the problem of describing galaxy rotation curves  {\it without dark matter} but asking for corrections to the Newtonian potential that could fit data and reproduce dynamics.
These corrections are not phenomenological but come out from the weak field limit of general relativistic theories of gravity that predict the existence of corrections (e.g. Yukawa-like corrections) to the Newtonian potential. The only exception  is GR where the action is chosen to be $R$, that is linear in the Ricci curvature scalar and does not contain corrections to the Newtonian potential in the weak field limit. Relaxing such a hypothesis, it is possible to show that {\it any analytic ETG presents Yukawa corrections in the weak-field limit} (see also \cite{Qua91} for a detailed caculation). From an astrophysical point of view, these corrections means that further scales have to be taken into account and that their effects  could be irrelevant at local scales as Solar System. 
With this scheme in mind, we will give a  summary of ETGs in  Sec. \ref{due} discussing also their conformal properties. In fact  any ETG can be conformally transformed to the Einstein one plus  scalar fields representing the further gravitational degrees of freedom. This feature is extremely important to select characteristic length scales (related to the effective masses of scalar fields) that could account for dynamics. In this sense, considering $f(R)$ gravity means to take into account an Einstein theory plus a scalar field; considering $f(R,\,\Box R)$-gravity means to assumes Einstein + two scalar fields and so on.  
The emergence of Yukawa-like corrections to the Newtonian potential is discussed in Sec. \ref{tre} where the weak-field limit of $f(R)$-gravity, the simplest ETG, is worked out. Here, $f(R)$ is a generic analytic  function of the Ricci curvature scalar $R$.
Furthermore,  we discuss the case   of $f(R,\phi)$-gravity, corresponding to $f(R,\Box R)$-gravity, i.e. Einstein plus two scalar fields, showing that a further free parameter in needed  to better model dynamics. 
Sec. \ref{cinque} is devoted to the rotation curves of galaxies. It is shown that the phenomenological Sanders potential, suitable to fit realistically observations, can be  reproduced by  the weak filed limit of $f(R,\phi)$-gravity.  Sec \ref{sei} is devoted to discussion and conclusions.

\section{Extended Gravity and Conformal trasformations}
\label{due}
 Higher-order and scalar-tensor gravities are examples of ETGs.  For a comprehensive discussion, see 
\cite{PRnostro,reviewodi,reviewodi1,reviewmauro,reviewvalerio,libro}. Essentially these theories can be characterized by  
 two main feature: the geometry can 
 non-minimally couple to some scalar 
field;  derivatives of the metric components of order higher  than second  
may appear.
 In the first case, we say that we have  
scalar-tensor  gravity, and in the second case we have  
higher-order theories. Combinations of non-minimally 
coupled and higher order terms can 
also emerge in effective 
Lagrangians, producing mixed higher order/scalar-tensor 
gravity. The physical foundation of such models can be found at fundamental  level by considering effective actions coming from quantum fields in curved space-times, string/M theory and so on  \cite{libro}. 
 A general class of higher-order-scalar-tensor theories in four dimensions is
 given by the effective action
 
 \begin{eqnarray} \label{V3.1}
 {\cal S}&=&\int d^{4}x\sqrt{-g}\left[f(R,\Box R,\Box^{2}R,\dots,\Box^kR,\phi)+\omega(\phi)
\phi_{; \alpha} \phi^{;\alpha}+ \mathcal{X} \mathcal{L}_m\right],
\end{eqnarray}
where $f$ is an unspecified function of curvature invariants and  scalar
field $\phi$ and $\mathcal{X}\,=\,8\pi G$\footnote{Here we use the convention
$c\,=\,1$.}. The convention for Ricci's tensor is
$R_{\mu\nu}={R^\sigma}_{\mu\sigma\nu}$, while for the Riemann
tensor is
${R^\alpha}_{\beta\mu\nu}=\Gamma^\alpha_{\beta\nu,\mu}+...$. The
affinities are the usual Christoffel symbols of the metric:
$\Gamma^\mu_{\alpha\beta}=\frac{1}{2}g^{\mu\sigma}(g_{\alpha\sigma,\beta}+g_{\beta\sigma,\alpha}
-g_{\alpha\beta,\sigma})$. The adopted signature is $(+---)$).
The term $\mathcal{L}_m$  is the minimally
coupled ordinary matter contribution, considered  as  a {\it perfect fluid}; $\omega(\phi)$ is a function of scalar field which specifies the theory. Actually its
 values can be $\omega(\phi) =\pm 1,0$ fixing the nature and the
 dynamics of the scalar field which can be a canonical scalar
 field, a phantom field or a field without dynamics (see
 \cite{valerio,odi2005,singularity} for details).
In the metric approach, the field equations are obtained by
varying (\ref{V3.1}) with respect to  $g_{\mu\nu}$.  By introducing the Einstein tensor $G_{\mu\nu}$ we get

\begin{eqnarray} \label{3.2cc}
\mathcal{G}\,G_{\mu\nu}\,=&&\mathcal{X}\,T_{\mu\nu}+\frac{f-{\cal G}R}{2}g_{\mu\nu}+
{\cal G}_{;\mu\nu}-g_{\mu\nu}\Box\mathcal{G}-\omega(\phi)\biggl(\phi_{;\mu}\phi_{;\nu}-
\frac{\phi_{;\alpha}\phi^{; \alpha}}{2}g_{\mu\nu}\biggr)\nonumber\\&&+\frac{1}{2}\sum_{i=1}^{k}\sum_{j=1}^{i}(g_{\mu\nu}
g^{\lambda\sigma}+g_\mu^{\,\,\,\lambda} g_\nu^{\,\,\,\sigma})(\Box^{j-i})_{;\sigma}
\times\left(\Box^{i-j}\frac{\partial f}{\partial \Box^{i}R}\right)_{;\lambda}-g_{\mu\nu}\left((\Box^{j-1}R)_{;\sigma}
\Box^{i-j}\frac{\partial f}{\partial \Box^{i}R}\right)^{;\sigma},
\end{eqnarray}
where we have introduced the quantity 

\begin{eqnarray} 
\label{3.4gg}
  {\cal G}\equiv\sum_{j=0}^{n}\Box^{j}\left(\frac{\partial f}{\partial \Box^{j} R}
\right)\,,\end{eqnarray}
the energy-momentum tensor of matter

\begin{eqnarray}\label{en_ten}
T_{\mu\nu}\,=\,-\frac{1}{\sqrt{-g}}\frac{\delta(\sqrt{-g}\,\mathcal{L}_m)}{\delta
g^{\mu\nu}}
\end{eqnarray}
and $\Box={{}_{;\sigma}}^{;\sigma}$ is the d'Alembert operator. The differential Eqs.(\ref{3.2cc}) are of order at most 
$(2k+4)$. The (possible) contribution of
a self-interaction potential $V(\phi)$ is contained in the definition of $f$.
By varying with respect to the scalar field $\phi$, we obtain the generalized
Klein-Gordon equation

\begin{eqnarray}\label{3.62}
2\,\omega(\phi)\,\Box\,\phi+\omega_{\phi}(\phi)\,\phi_{;\alpha}\phi^{;\alpha}-f_{\phi}\,=\,0\,,
\end{eqnarray}
where $f_\phi\,=\,\frac{df}{d\phi}$ and
$\omega_\phi(\phi)\,=\,\frac{d\omega(\phi)}{d\phi}$. Several interesting cases can be worked out starting from the action (\ref{V3.1}). Below, we give some significant examples that will result useful for the astrophysical applications of this paper.

\subsection{The case of $f(R)$-gravity}
The simplest extension of GR is achieved assuming, 
 
 \begin{eqnarray}\label{fr}
R\rightarrow\,f(R)\,,\qquad \omega(\phi)\,=\,0\,
\end{eqnarray} 
and the action (\ref{V3.1}) becomes 

\begin{equation} \label{s1_fR} 
{\cal S}= \int  d^{4}x\sqrt{-g}\left[f(R)+\mathcal{X}\mathcal{L}_m\right]\,.
\end{equation}
Then the field equations (\ref{3.2cc}) become 

\begin{eqnarray}\label{VAR12.34}
f_R\,G_{\mu\nu}\,=\,\mathcal{X}T_{\mu\nu}+\frac{f-f_R R}{2}g_{\mu\nu}+
{f_R}_{;\mu\nu} - g_{\mu\nu}\Box f_R\,\equiv\,\mathcal{X}T_{\mu\nu}+f_RT^{f(R)}_{\mu\nu}
\end{eqnarray}  
where ${\displaystyle f_R\,=\,\frac{df}{dR}}$. The gravitational contribution due to higher-order terms can
be  reinterpreted as a stress-energy tensor contribution $T^{f(R)}_{\mu\nu}$.
This means that additional and higher-order terms in the
gravitational action act, in principle, as a stress-energy tensor,
related to the  form of $f$.
In the case of GR,   $T^{f(R)}_{\mu\nu}$ identically vanishes while the
standard, minimal coupling is recovered for the matter
contribution.

The peculiar behavior of $f(R)\,=\,R$ is  due to the
particular form of the Lagrangian itself which, even though it is
a second order Lagrangian, can be non-covariantly rewritten as the
sum of a first order  Lagrangian plus a pure divergence term. The
Hilbert-Einstein Lagrangian can be in fact recast as follows:
\begin{eqnarray}
L_{HE}&=& {\cal L}_{HE} \sqrt{-g}=\Big[ p^{\alpha \beta}
(\Gamma^{\rho}_{\alpha \sigma} \Gamma^{\sigma}_{\rho
\beta}-\Gamma^{\rho}_{\rho \sigma} \Gamma^{\sigma}_{\alpha
\beta})+ \nabla_\sigma (p^{\alpha \beta} {u^{\sigma}}_{\alpha
\beta}) \Big]\,,\nonumber\\
\end{eqnarray}
\noindent where:
\begin{equation}
 p^{\alpha \beta} =\sqrt{-g}  g^{\alpha \beta} = \frac{\partial {\cal{L}}}{\partial R_{\alpha \beta}}\,,
\end{equation}
$\Gamma$ is the Levi-Civita connection of $g$ and
$u^{\sigma}_{\alpha \beta}$ is a quantity constructed out with the
variation of $\Gamma$ \cite{weinberg}. Since $u^{\sigma}_{\alpha
\beta}$ is not a tensor, the above expression is not covariant;
however  standard procedures  can be used to recast covariance \cite{libro}. This clearly shows that
the field equations has to be of second order  and the
Hilbert-Einstein Lagrangian is thus degenerate.

\subsection{The case of scalar-tensor gravity}

From the action (\ref{V3.1}), it is possible to obtain another
interesting case by choosing

\begin{eqnarray}
f\,=\,F(\phi)R+V(\phi)\,,\qquad \omega(\phi)\,=\,1/2\,,
\end{eqnarray}
then

\begin{equation} \label{s1} 
{\cal S}= \int  d^{4}x\sqrt{-g}\left[F(\phi) R+V(\phi)+\frac{\phi_{;\alpha}\phi^{;\alpha}}{2}+\mathcal{X}\mathcal{L}_m\right]\,,
\end{equation}
where $V(\phi)$ and $F(\phi)$ are generic functions describing respectively the
potential and the coupling of a scalar field $\phi$.  The
Brans-Dicke theory of gravity is a particular case of the action
(\ref{s1}) for $V(\phi)$\,=\,0 \cite{libro}. The variation with respect to $g_{\mu\nu}$ gives now the second-order field equations (particular form of field equations (\ref{3.2cc}))

\begin{equation} \label{s2} 
F(\phi)\,G_{\mu\nu}\,=\,\mathcal{X}T_{\mu\nu}+\frac{V(\phi)}{2}g_{\mu\nu} +F(\phi)_{;\mu\nu}- g_{\mu\nu} \Box F(\phi)-\frac{1}{2}\biggl(\phi_{;\mu}\phi_{;\nu}-
\frac{\phi_{;\alpha}\phi^{; \alpha}}{2}g_{\mu\nu}\biggr)\,\equiv\,\mathcal{X}T_{\mu\nu}+F(\phi)T^{(\phi)}_{\mu\nu}
\end{equation} 
 where $T^{(\phi)}_{\mu\nu}$ is the energy-momentum
tensor relative to the scalar field $\phi$. The variation with respect
to $\phi$ provides the Klein - Gordon equation, {\it i.e.} the field
equation for the scalar field

\begin{equation} \label{s5}
\Box \phi-F_{\phi}(\phi)R-V_{\phi}(\phi)= 0\,,
\end{equation}
where $\displaystyle{F_{\phi}(\phi)= \frac{dF(\phi)}{d\phi}}$, $\displaystyle{V_{\phi}(\phi)= \frac{dV(\phi)}{d\phi}}$. This last equation
is equivalent to the Bianchi contracted identity \cite{cqg}.

\subsection{Conformal transformations}

These models, and, in general, any theory of the class (\ref{V3.1}), can be conformally reduced to the Einstein theory plus scalar fields.
Conformal transformations are  mathematical tools  very useful in  ETGs
 in order to disentangle the further gravitational degrees of freedom coming from general actions \cite{libro,MagnanoSokolowski94,FGN98,FaraoniNadeau}.  The idea is to perform a conformal rescaling of the   
space-time metric $g_{\mu\nu} \rightarrow \tilde{g}_{\mu\nu}$. 
Often a  scalar field is present in the theory and the metric 
rescaling  is accompanied by a (nonlinear) redefinition of this 
field $\phi  \rightarrow \tilde{\phi}$.  New dynamical variables
$ \left\{ \tilde{g}_{\mu\nu} , \tilde{\phi} \right\}$ are thus 
obtained. The scalar field redefinition serves the purpose of 
casting the kinetic energy density of this field in a canonical 
form. The new set of variables $\left\{\tilde{g}_{\mu\nu}, 
\tilde{\phi} \right\}$ is called the {\em Einstein conformal 
frame}, while $\left\{ g_{\mu\nu}, \phi 
\right\}$ constitute the {\em 
Jordan frame}. When a scalar degree of 
freedom $\phi$ is present 
in  the theory, as in scalar tensor 
or $f(R)$ gravity, it generates the 
transformation to  the Einstein frame  in 
the sense that the 
rescaling is completely determined by a function of $\phi$. In 
principle, infinitely 
many conformal frames could be introduced, giving rise to as many 
representations of the theory. 

Let the pair  $\{{\cal M}, g_{\mu\nu}\}$ be  a space-time, with ${\cal M}$ a smooth 
manifold of  dimension ~$n \geq 2$ and $g_{\mu\nu} $ a 
(pseudo)-Riemannian metric on ${\cal M}$. The point-dependent rescaling of the 
metric tensor

\begin{eqnarray}   \label{cft33}
g_{\mu\nu} \longrightarrow \tilde{g}_{\mu\nu}=\Omega^2  
g_{\mu\nu} \, ,
\end{eqnarray}
where the {\em conformal factor}
$\Omega$ is a nowhere
vanishing, regular function, is called a {\em Weyl} or {\em conformal} 
transformation. Due to this metric rescaling, the 
lengths of space-like and time-like intervals and the norms of 
space-like and time-like vectors are changed, while  null vectors 
and null intervals of the metric $g_{\mu\nu}$ remain 
null in the rescaled metric $\tilde{g}_{\mu\nu}$.  The light cones 
are left unchanged by the transformation (\ref{cft33}) 
and the  space-times  $\{{\cal M}, g_{\mu\nu}\}$ and  $\{{\cal M}, 
\tilde{g}_{\mu\nu}\} $ 
exhibit the same causal structure; the converse is also true 
\cite{Wald84}. A vector that is time-like, space-like,
or null with respect to the  metric $g_{\mu\nu}$ has the same 
character with respect to $\tilde{g}_{\mu\nu}$, and 
{\em vice-versa}.

Conformal invariance corresponds to the absence of a characteristic length (or mass) 
scale in the physics. In general, the effective potential $V(\phi)$  coming from conformal transformations 
contains dimensional parameters  (such as a mass $m$, that is  a further "characteristic gravitational length"). 
This means that the further degrees of freedom coming from ETGs give rise to features that could play a fundamental role in the dynamics of astrophysical structures. In what follows, we will see that these further gravitational lengths could solve, in principle, the dark matter problem.

\subsection{Conformal transformations and higher-order gravity}

Performing the conformal transformation for $f(R)$-gravity with $\Omega^2\,=\,f_R$ we have 

\begin{eqnarray} \label{h7}
\int d^4x\sqrt{-g}[f(R)+\mathcal{X}\mathcal{L}_m]\,=\,\int d^4x\sqrt{-\tilde{g}} \left(\tilde{R}+
W(\tilde{\phi})-\frac{\tilde{\phi}_{;\alpha}\tilde{\phi}^{;\alpha}}{2}+\mathcal{X}\tilde{\mathcal{L}}_m\right)
\end{eqnarray}
where $\tilde{\phi}\,=\,\sqrt{3}\,\ln f_R$ while the potential $W$ and the nonminimally coupled lagrangian of ordinary matter $\tilde{\mathcal{L}}_m$ are given by

\begin{eqnarray}\label{transconfTS}
&&W(\tilde{\phi})\,=\,e^{-2\,\tilde{\phi}/\sqrt{3}}\,V(e^{\tilde{\phi}/\sqrt{3}})\nonumber\\\\
&&\tilde{\mathcal{L}}_m\,=\,e^{-2\,\tilde{\phi}/\sqrt{3}}\,\mathcal{L}_m\biggl(e^{-\tilde{\phi}/\sqrt{3}}\tilde{g}_{\rho\sigma}\biggr)
\nonumber
\end{eqnarray}
The function $V$ is defined by the analogy between the $f(R)$-gravity and the scalar-tensor gravity (the so-called O'Hanlon lagrangian)

\begin{eqnarray}\label{h4_a}
V(\phi)\,=\,f(R)-R\,f_R(R)
\end{eqnarray}
where $\phi\,=\,f_R$. The field equations in standard form are given in the Einstein frame as follow

\begin{eqnarray}\label{h4}
\tilde{G}_{\mu\nu}\,=\,\mathcal{X}\,\tilde{T}_{\mu\nu}+\frac{W(\tilde{\phi})}{2}\,\tilde{g}_{\mu\nu}+\frac{1}{2}\biggl(\tilde{\phi}_{;\mu}\tilde{\phi}_{;\nu}-
\frac{\tilde{\phi}_{;\alpha}\tilde{\phi}^{;\alpha}}{2}\tilde{g}_{\mu\nu}\biggr)
\end{eqnarray}

\begin{eqnarray}\label{h4_sf}
\tilde{\Box}\,\tilde{\phi}+W_{\tilde{\phi}}(\tilde{\phi})\,=\,
-\mathcal{X}\,\frac{\delta\tilde{\mathcal{L}}_m}{\delta\tilde{\phi}}
\end{eqnarray}
However, the problem is completely solved if $\phi\,=\,f_R$ can be analytically inverted. In summary, a fourth-order theory is conformally equivalent to the standard
second-order Einstein theory plus a scalar field (see also
\cite{francaviglia,ordsup}).

If the theory is higher than fourth order, we have Lagrangian
densities of the form \cite{buchdahl,gottloeber,sixth},

\begin{eqnarray}\label{h10}
\mathcal{L}\,=\,\mathcal{L}(R,\,\Box R,\,\dots,\,\Box^{k} R)\,.
\end{eqnarray}
Every $\Box$ operator introduces two further terms of derivation
into the field equations. For example a theory like

\begin{eqnarray}\label{h11}
\mathcal{L}\,=\,R\,\Box R\,,
\end{eqnarray}
is a sixth-order theory, and the above approach can be pursued considering a conformal
factor of the form

\begin{eqnarray}\label{h12}
\Omega^2\,=\,\frac{\partial\mathcal{L}}{\partial R} +\Box\frac{\partial\mathcal{L}}{\partial \Box R}\,.
\end{eqnarray}
In general,  increasing two orders of derivation in the field
equations (\emph{i.e.} every term $\Box R$), corresponds to add a scalar
field in the conformally transformed frame \cite{gottloeber}. A
sixth-order theory can be reduced to an Einstein theory with two
minimally coupled scalar fields; a $2n$-order theory can be, in
principle, reduced to an Einstein theory + $(n-1)$-scalar fields.
On the other hand, these considerations can be directly
generalized to higher-order-scalar-tensor theories in any number
of dimensions as shown in \cite{libro}.
With these considerations in mind, we can easily say that a higher order theory like $f(R,\Box R)$ is dynamically equivalent to  $f(R,\phi)$.
This feature, as we will show,  gives the minimal ingredients to reproduce the rotation curves of galaxies since two Yukawa like corrections come out. For a detailed derivation see \cite{Qua91}.

\section{Yukawa-like corrections to the gravitational potential}
\label{tre}

In order to deal with standard self-gravitating systems, any theory of gravity has to be developed to its Newtonian or post-Newtonian limit depending on the order of approximation of the theory in terms of power of velocity $v^2$ \cite{cqg,stabile}. The paradigm of the Newtonian limit starts from the development of the
metric tensor (and of all additional quantities in the theory) with respect to the dimensionless velocity\footnote{The velocity $v$ is expressed in unit of light speed.} $v$ of the moving massive bodies embedded in the gravitational potential. The perturbative development takes  only first term of $0,0$- and $i,j$-component of metric tensor $g_{\mu\nu}$ (for details, see \cite{PRD1,PRD1_2}). The  metric assumes the following form

\begin{eqnarray}\label{me0}
{ds}^2\,=\,(1+2\Phi)\,dt^2-(1-2\Psi)\,\delta_{ij}dx^idx^j
\end{eqnarray}
where the gravitational potentials $\Phi,\, \Psi\,<\,1$ are proportional to $v^2$. The adopted set of coordinates\footnote{The Greek index runs from $0$ to $3$; the Latin index runs from $1$ to $3$.},  the so-called {\it isotropic coordinates}, is $x^\mu\,=\,(t,\textbf{x})=\,(t,x^1,x^2,x^3)$. The Ricci scalar is approximated as $R\,=\,R^{(1)}\,+\,R^{(2)}\,+\,\dots$ where $R^{(1)}$ is proportional to $\Phi$, and $\Psi$, while $R^{(2)}$ is proportional to $\Phi^2$, $\Psi^2$ and $\Phi\Psi$.

Here we show as a general gravitational potential, with a Yukawa correction, can be
obtained in the Newtonian limit of any analytic $f(R)$-gravity
model. From a phenomenological point of view, this correction
allows to consider as viable this kind of models even at small
distances, provided that the Yukawa correction turns out to be
not relevant in this approximation as in the so called "chameleon
mechanism" \cite{chameleon}.

\subsection{Yukawa-like corrections in  $f(R)$-gravity}

Starting from the action (\ref{V3.1}) for the case $f(R,\Box R,\Box^{2}R,\dots,\Box^kR,\phi)+\omega(\phi)
\phi_{; \alpha} \phi^{;\alpha}$ reduced to $f(R)$, the field equations are (\ref{VAR12.34}).
In principle, the following analysis can be developed for any ETGs.  Let us now start with the  $f(R)$ case.
As discussed in \cite{PRnostro,arturosferi,arturonoether}, we can deal with the
Newtonian limit of $f(R)$ gravity
adopting the spherical symmetry. By introducing the radial coordinate $r\,=\,|\textbf{x}|$ the metric (\ref{me0}) can be recast as follows

\begin{eqnarray}\label{me}
ds^2\,=\,[1+g^{(1)}_{tt}(t,r)]\,dt^2-[1-g^{(1)}_{rr}(t,r)]\,dr^2-r^2\,d\Omega\,,
\end{eqnarray}
where $d\Omega\,=\,d\theta^2+\sin^2\theta\,d\phi^2$ is the angular distance, $(t,r,\theta,\phi)$ are  standard coordinates. Since we want to obtain the most general result, we do not provide any specific form for the $f(R)$. We assume, however, analytic Taylor expandable $f(R)$ functions with respect to the value $R\,=\,0$ (Minkowskian background):

\begin{eqnarray}\label{sertay}
f(R)\,=\,\sum_{n\,=\,0}^{\infty}\frac{f^{(n)}(0)}{n!}\,R^n\,=\,
f_0+f_1R+\frac{f_2}{2}R^2+\frac{f_3}{6}R^3+...
\end{eqnarray}
In order to obtain the weak field  approximation, one has to
insert  expansions (\ref{me}) and (\ref{sertay}) into
field Eqs. (\ref{VAR12.34})  and expand the system up
to the orders ${\mathcal O}(0)$ and ${\mathcal O}(1)$. This approach provides
general results and specific (analytic) theories are selected by
the coefficients $f_i$ in Eq.(\ref{sertay}). It is worth noticing
that, at the order ${\mathcal O}(0)$, the field equations give the condition
$f_0 =0$ and then the solutions at further orders do not depend on
this parameter as we will show below. If we now consider the
${\mathcal O}(1)$ - order approximation, the field equations  in
vacuum ($T_{\mu\nu}\,=\,0$), results to be

\begin{eqnarray}\label{eq2}
&&f_1rR^{(1)}-2f_1g^{(1)}_{tt,r}+4f_2R^{(1)}_{,r}-f_1rg^{(1)}_{tt,rr}+2f_2rR^{(1)}=0\,,\nonumber
\\\nonumber\\
&&f_1rR^{(1)}-2f_1g^{(1)}_{rr,r}+4f_2R^{(1)}_{,r}-f_1rg^{(1)}_{tt,rr}=0\,,\nonumber
\\\nonumber\\
&&2f_1g^{(1)}_{rr}-r\left[f_1rR^{(1)}-f_1g^{(1)}_{tt,r}-f_1g^{(1)}_{rr,r}+2f_2R^{(1)}_{,r}+2f_2rR^{(1)}_{,rr}\right]=0\,,
\\\nonumber\\
&&f_1rR^{(1)}+3f_2\left[2R^{(1)}_{,r}+rR^{(1)}_{,rr}\right]=0\,,\nonumber
\\\nonumber\\
&&2g^{(1)}_{rr}+r\left[2g^{(1)}_{tt,r}-rR^{(1)}+2g^{(1)}_{rr,r}+rg^{(1)}_{tt,rr}\right]=0\,.\nonumber
\end{eqnarray}
It is evident that the trace equation (the fourth in the system
(\ref{eq2})), provides a differential equation with respect to the
Ricci scalar which allows to solve  exactly the system (\ref{eq2})
at ${\mathcal O}(1)$ - order. Finally, one gets the general solution\,:

\begin{eqnarray}\label{sol}
&&g^{(1)}_{tt}\,=\,\delta_0-\frac{Y}{f_1r}+\frac{\delta_1(t)e^{-mr}}{3m^2
r}+\frac{\delta_2(t)e^{mr}}{6m^3r}\nonumber
\\\nonumber\\
&&g^{(1)}_{rr}\,=\,-\frac{Y}{f_1r}-\frac{\delta_1(t)[1+mr]e^{-mr}}{3m^2
r}-\frac{\delta_2(t)[1-mr]e^{mr}}{6m^3r}
\\\nonumber\\
&&R^{(1)}\,=\,\frac{\delta_1(t)\,e^{-mr}}{r}+\frac{\delta_2(t)e^{mr}}{2m
r}\nonumber
\end{eqnarray}
where $m^2\doteq-\frac{f_1}{3f_2}$, $\delta_0$ and $Y$ are arbitrary constants, while $\delta_1(t)$ and $\delta_2(t)$
are arbitrary time functions. When we consider the limit $f(R)\rightarrow R$ then $m\rightarrow\infty$ and $f_1\,\rightarrow\,1$, in the case
of a point-like source of mass $M$, we recover the standard
Schwarzschild solution if we set $\delta_0\,=\,0$ and $Y\,=\,2GM$.  Let us notice that the
integration constant $\delta_0$ is  dimensionless, while the two functions $\delta_1(t)$ and $\delta_2(t)$ have
respectively the dimensions of $length^{-1}$ and $length^{-2}$.  These functions are completely arbitrary since the
differential equation system (\ref{eq2}) contains only spatial
derivatives and can be fixed to  constant values. Besides, the condition $\delta_0\,=\,0$ is avaible
since it represents an unessential additive quantity for the potential.

The solutions (\ref{sol}) are valid if $m^2\,>\,0$ \emph{i.e.} $f_1$ and $f_2$ are assumed to have different signs in Eq. (\ref{sertay}). If the algebraic signs are equal we find an oscillating solution where the correction term to the newtonian component ($\propto\,1/r$) is proportional to the $(\cos m r+\sin mr)/r$  \cite{PRD1}. In this paper we consider only the  correction Yukawa-like.

It is possible now, to write  the general solution of the problem considering
the previous expressions (\ref{sol}). In order to match at infinity
the Minkowskian prescription for the metric, one can discard the
Yukawa growing mode  in (\ref{sol}) and then we obtain\,:

\begin{eqnarray}\label{mesol}
&&ds^2\,=\,\biggl[1-\frac{2GM}{f_1r}+\frac{\delta_1(t)e^{-mr}}{3m^2
r}\biggr]dt^2- \biggl[1+\frac{2GM}{f_1r}+\frac{\delta_1(t)[1+mr]e^{-mr}}{3m^2
r}\biggr]dr^2-r^2d\Omega\,,\nonumber\\\\
&&R\,=\,\frac{\delta_1(t)e^{-mr}}{r}\,.\nonumber
\end{eqnarray}

At this point, one can provide the solution in terms of 
gravitational potentials.  The first of (\ref{sol}) gives the first order solution in term of the metric expansion (see the
definition (\ref{me})). This term coincides with the
gravitational potential at the Newton order. In particular, since
 $g_{tt}\,=\,1+2\Phi_{grav}\,=\,1+g_{tt}^{(1)}$, the 
gravitational potential of $f(R)$-gravity, analytic in the Ricci
scalar $R$, is

\begin{eqnarray}\label{gravpot}
\Phi_{grav}\,=\,-\left(\frac{GM}{f_1r}-\frac{\delta_1(t)e^{-mr}}{6\,m^2
r}\right)\,.
\end{eqnarray}
This general result means that the standard Newton potential is
achieved only in the particular case $f(R)=R$ while it is not so
for analytic $f(R)$ models up to exceptions of measure zero. Specifically all models with $f_1/f_2\,>\,0$ are excluded by hand. Eq.(\ref{gravpot}) deserves
some comments. The parameters $f_1$, $m$ and the function
$\delta_1(t)$ represent the deviations with respect the standard
Newton potential. To test these theories of gravity inside the
Solar System, we need to compare such quantities with respect to
the current experiments, or, in other words, Solar System
constraints  should be evaded fixing such parameters \cite{chameleon}. On the other
hand, these parameters could acquire  non-trivial values (e.g.
$f_1\neq 1,\,\delta_1(t)\neq 0,\,m\,<\,\infty$) at   scales different
from the Solar System ones.
Since the parameter $m$ can be  related to an effective length $L^{-1}$,
Eq. (\ref{gravpot}) can be recast as 

\begin{eqnarray}\label{gravpot1}
\Phi_{grav}\,=\,-\left(\frac{G M}{1+\delta}\right)\frac{1+\delta\,e^{-r/L}}{r}\,,
\end{eqnarray}
where the first term is the Newtonian-like part of the potential
for a point-like mass ${\displaystyle \frac{M}{1+\delta}}$ and the second
term is a modification of  gravity including a new scale
length, $L$ associated to the coefficients of the Taylor expansion.
If $\delta\,=\,0$ the Newtonian potential and the standard gravitational coupling are  recovered. 
Comparing Eqs. (\ref{gravpot}) and (\ref{gravpot1}), we assumed 
$f_1\,=\,1+\delta$ and ${\displaystyle \delta_1(t)\,=\,-\frac{6\,GM}{L^2}\left(\frac{\delta}{1+\delta}\right)}$
where $\delta$ can be chosen quasi-constant.  
Under this assumption, the scale length $L$ could naturally arise and reproduce several phenomena that range from Solar System to large scale structure
 \cite{annalen}. Understanding 
on which scales the modifications to GR are working or what is the weight of corrections 
to Newton potential is a crucial point that could confirm or rule out these extended approaches to gravitational interaction. 

\subsection{Yukawa-like corrections  in  $f(R,\phi)$-gravity}

A further step is to analyze the Newtonian limit starting from the action (\ref{V3.1}) and considering a generic function of Ricci scalar and scalar field. Then the action becomes

\begin{eqnarray}\label{HOGaction}
\mathcal{A}=\int d^{4}x\sqrt{-g}\biggl[f(R,\phi)+\omega(\phi)\,\phi_{;\alpha}\,\phi^{;\alpha}+\mathcal{X}\mathcal{L}_m\biggr]
\end{eqnarray}
The field equations are obtained from (\ref{3.2cc}) by setting $f(R,\Box R,\Box^{2}R,\dots,\Box^kR,\phi)\,\rightarrow\,f(R,\phi)$.  As discussed in Sec. II, this case can be considered from a purely geometric point of view assuming $f(R,\Box R)$ theories where the terms $\Box R$ gives a further scalar field contribution \cite{Qua91}. We get

\begin{eqnarray}
\label{fieldequationHOG}
&&f_RR_{\mu\nu}-\frac{f+\omega(\phi)\,\phi_{;\alpha}\,\phi^{;\alpha}}{2}\,g_{\mu\nu}+\omega(\phi)\,\phi_{;\mu}\,\phi_{;\nu}-f_{R;\mu\nu}+g_{\mu\nu}\Box\,
f_R\,=\,\mathcal{X}\,T_{\mu\nu}\nonumber\\\\
&&2\,\omega(\phi)\,\Box\,\phi+\omega_{\phi}(\phi)\,\phi_{;\alpha}\phi^{;\alpha}-f_{\phi}\,=\,0\nonumber
\end{eqnarray}
A further equation is the trace of field equation with respect to the metrci tensor $g_{\mu\nu}$
\begin{eqnarray}\label{trace}
f_R\,R-2f-\omega(\phi)\,\phi_{;\alpha}\phi^{;\alpha}+3\,\Box\,f_R\,=\,\mathcal{X}\,T
\end{eqnarray}
where $T\,=\,T^{\sigma}_{\,\,\,\,\,\sigma}$ is the trace of energy-momentum tensor.

Let us consider a point-like source with mass $M$. The energy-momentum tensor is 
\begin{eqnarray}\label{emtensor}
T_{\mu\nu}\,=\,\rho\,u_\mu
u_\nu\,,\,\,\,\,\,\,\,\,\,\,T\,=\,\rho
\end{eqnarray}
where $\rho$ is the mass density and $u_\mu$ satisfies the condition $g^{00}{u_0}^2\,=\,1$ and $u_i\,=\,0$.  Here, we are not interested  to the internal structure.
It is possible to analyze the problem in the more general case by using the isotropic coordinates $(t,x^1,x^2,x^3)$, then the metric is expressed as in Eq. (\ref{me0}).
In this framework,  also the scalar field $\phi$ is approximated as the Ricci scalar. In particular we get $\phi\,=\,\phi^{(0)}\,+\,\phi^{(1)}\,+\,\phi^{(2)}\,+\dots$ and the function $f(R,\phi)$ with its partial derivatives ($f_R$, $f_{RR}$, $f_{\phi}$, $f_{\phi\phi}$ anf $f_{\phi R}$) and $\omega(\phi)$ can be substituted by their corresponding Taylor series. In the case of $f(R,\phi)$, we have

\begin{eqnarray}
f(R,\phi)\,\sim\,f(0,\phi^{(0)})+f_R(0,\phi^{(0)})\,R^{(1)}+f_\phi(0,\phi^{(0)})\,\phi^{(1)}+\dots
\end{eqnarray}
and analogous relations for the derivatives are obtained. From the lowest order of field Eqs. (\ref{fieldequationHOG}) we have

\begin{eqnarray}\label{PPN-field-equation-general-theory-fR-O0}
f(0,\,\phi^{(0)})\,=\,0\,,\,\,\,\,\,\,\,\,\,\,f_{\phi}(0,\phi^{(0)})\,=\,0
\end{eqnarray}
and also in this modified fourth order gravity a missing cosmological component in the action (1) implies that the space-time is asymptotically Minkowskian (the same outcome of previous section); moreover the ground value of scalar field $\phi$ must be a stationary point of potential. In the Newtonian limit,  we have

\begin{eqnarray}
\label{NL-fieldequationHOG}
&&\triangle\biggl[\Phi-\frac{f_{RR}(0,\phi^{(0)})}{f_R(0,\phi^{(0)})}\,R^{(1)}-\frac{f_{R\phi}(0,\phi^{(0)})}{f_R(0,\phi^{(0)})}\,\phi^{(1)}\biggr]-\frac{R^{(1)}}{2}\,=\,\frac{\mathcal{X}\,\rho}{f_R(0,\phi^{(0)})}\nonumber\\\nonumber\\
&&\biggl\{\triangle\biggl[\Psi+\frac{f_{RR}(0,\phi^{(0)})}{f_R(0,\phi^{(0)})}\,R^{(1)}+\frac{f_{R\phi}(0,\phi^{(0)})}{f_R(0,\phi^{(0)})}\,\phi^{(1)}\biggr]+\frac{R^{(1)}}{2}\biggr\}\delta_{ij}+\nonumber\\\nonumber\\
&&\,\,\,\,\,\,\,\,\,\,\,\,\,\,\,\,\,\,\,\,\,\,\,\,\,\,\,\,\,\,\,\,\,\,\,\,\,\,\,\,\,\,\,\,\,\,\,\,\,\,\,\,\,\,\,\,\,\,\,\,\,\,\,\,\,\,\,\,\,\,\,\,\,\,\,\,\,\,\,\,\,\,\,\,+\biggr\{\Psi-\Phi-\frac{f_{RR}(0,\phi^{(0)})}{f_R(0,\phi^{(0)})}R^{(1)}-\frac{f_{R\phi}(0,\phi^{(0)})}{f_R(0,\phi^{(0)})}\phi^{(1)}\biggr\}_{,ij}\,=\,0\\\nonumber\\
&&\triangle\phi^{(1)}+\frac{f_{\phi\phi}(0,\phi^{(0)})}{2\,\omega(\phi^{(0)})}\,\phi^{(1)}\,=\,-\frac{f_{R\phi}(0,\phi^{(0)})}{2\,\omega(\phi^{(0)})}\,R^{(1)}\nonumber\\\nonumber\\
&&\triangle R^{(1)}+\frac{f_R(0,\phi^{(0)})\,R^{(1)}}{3\,f_{RR}(0,\phi^{(0)})}\,=\,-\frac{\mathcal{X}\,\rho}{3\,f_{RR}(0,\phi^{(0)})}-\frac{f_{R,\phi}(0,\phi^{(0)})}{f_{RR}(0,\phi^{(0)})}\,\triangle\phi^{(1)}\nonumber
\end{eqnarray}
where $\triangle$ is the Laplacian in the flat space. The last equation in (\ref{NL-fieldequationHOG}) is the trace coming from  Eq. (\ref{trace}). These equations are not simply the merging of field equations of $f(R)$-gravity and a further massive scalar field, but  are due to the fact that the model $f(R,\phi)$ generates a coupled system of equations with respect to Ricci scalar $R$ and scalar field $\phi$. By supposing that $f_{\phi\phi}\,\neq\,0$ and obviosuly $f_{RR}\,\neq\,0$, we can introduce the two characteristic length scales

\begin{eqnarray}\label{mass_definition}
{m_R}^2\,\doteq\,-\frac{f_R(0,\phi^{(0)})}{3f_{RR}(0,\phi^{(0)})},\,\,\,\,\,\,\,\,
{m_\phi}^2\,\doteq\,-\frac{f_{\phi\phi}(0,\phi^{(0)})}{2\,\omega(\phi^{(0)})}
\end{eqnarray}
where the two masses are assumed to be real and this gives further restrictions to the set of viable models.
The gravitational potentials $\Phi$ and $\Psi$ are given by\footnote{The potential $\Psi$ can be found also as  $\Psi(\textbf{x})\,=\,\frac{1}{8\pi}\int
d^3\textbf{x}'\frac{R^{(1)}(\textbf{x}')}{|\textbf{x}-
\textbf{x}'|}+\frac{R^{(1)}(\textbf{x})}{3{m_R}^2}-\frac{f_{R,\phi}(0,\phi^{(0)})}{f_R(0,\phi^{(0)})}\,\phi^{(1)}(\textbf{x})$.}

\begin{eqnarray}\label{new_sol}
\Phi(\mathbf{x})\,&=&\,-\frac{\mathcal{X}}{4\pi\,f_R(0,\phi^{(0)})}\int
d^3\textbf{x}'\frac{\rho(\textbf{x}')}{|\textbf{x}-
\textbf{x}'|}-\frac{1}{8\pi}\int
d^3\textbf{x}'\frac{R^{(1)}(\textbf{x}')}{|\textbf{x}-
\textbf{x}'|}-\frac{R^{(1)}(\textbf{x})}{3{m_R}^2}+\frac{f_{R,\phi}(0,\phi^{(0)})}{f_R(0,\phi^{(0)})}\,\phi^{(1)}(\textbf{x})\nonumber\\\\
\Psi(\mathbf{x})\,&=&\,\Phi(\textbf{x})-\frac{R^{(1)}(\textbf{x})}{3{m_R}^2}+\frac{f_{R,\phi}(0,\phi^{(0)})}{f_R(0,\phi^{(0)})}\,\phi^{(1)}(\textbf{x})\nonumber\end{eqnarray}
while, for the Ricci scalar and the scalar field,  we have the coupled system of equations

\begin{eqnarray}
\label{coupledsyst}
&&\biggl[\triangle-{m_\phi}^2\biggr]\phi^{(1)}\,=\,-\frac{f_{R\phi}(0,\phi^{(0)})}{2\,\omega(\phi^{(0)})}\,R^{(1)}\nonumber\\\\
&&\biggl[\triangle-{m_R}^2\biggr]R^{(1)}\,=\,\frac{{m_R}^2\,\mathcal{X}\,\rho}{f_R(0,\phi^{(0)})}+\frac{3\,{m_R}^2\,f_{R\phi}(0,\phi^{(0)})}{f_R(0,\phi^{(0)})}\,\triangle\phi^{(1)}\nonumber
\end{eqnarray}
The definition of ${m_R}^2$ is the generalization of $m^2$ in the case of pure $f(R)$-gravity. By using the Fourier transformation the system (\ref{coupledsyst}) has the following solutions

\begin{eqnarray}
\label{coupledsyst_sol}
&&\phi^{(1)}(\textbf{x})\,=\,-\frac{{m_R}^2\,f_{R\phi}(0,\phi^{(0)})\,\mathcal{X}}{2\,\omega(\phi^{(0)})\,f_R(0,\phi^{(0)})}\int\frac{d^3\textbf{k}}{(2\pi)^{3/2}}\frac{\tilde{\rho}(\textbf{k})\,e^{i\textbf{k}\cdot\textbf{x}}}{(\textbf{k}^2+{k_1}^2)(\textbf{k}^2+{k_2}^2)}\nonumber\\\\
&&R^{(1)}(\textbf{x})\,=\,-\frac{{m_R}^2\,\mathcal{X}}{f_R(0,\phi^{(0)})}\int\frac{d^3\textbf{k}}{(2\pi)^{3/2}}\frac{\tilde{\rho}(\textbf{k})\,(\textbf{k}^2+{m_\phi}^2)\,e^{i\textbf{k}\cdot\textbf{x}}}{(\textbf{k}^2+{k_1}^2)(\textbf{k}^2+{k_2}^2)}\nonumber
\end{eqnarray}
where

\begin{eqnarray}
2\,{k_{1,2}}^2\,=\,{m_R}^2+{m_\phi}^2-\frac{3{f_{R\phi}(0,\phi^{(0)})}^2{m_R}^2}{2\,\omega(\phi^{(0)})\,f_R(0,\phi^{(0)})}\pm\sqrt{\biggl[{m_R}^2+{m_\phi}^2-\frac{3{f_{R\phi}(0,\phi^{(0)})\,}^2{m_R}^2}{2\,\omega(\phi^{(0)})f_R(0,\phi^{(0)})}\biggr]^2-4{m_R}^2{m_\phi}^2}
\end{eqnarray}
In order to understand the relevant physical consequences of solutions (\ref{new_sol}),  it is sufficient to analyze the point-like source framework. Then, if we consider $\tilde{\rho}(\textbf{k})\,=\,M/(2\pi)^{3/2}$, where $M$ is the mass, and ${k_{1,2}}^2\,>\,0$ Eqs. (\ref{coupledsyst_sol}) become

\begin{eqnarray}
\label{coupledsyst_sol_point}
&&\phi^{(1)}(\textbf{x})\,=\,
\frac{f_{R\phi}(0,\phi^{(0)})}{2\,\omega(\phi^{(0)})\,f_R(0,\phi^{(0)})}\frac{r_g}{|\textbf{x}|}\frac{e^{-m_R\tilde{k}_1\,|\textbf{x}|}-e^{-m_R\tilde{k}_2\,|\textbf{x}|}}{{\tilde{k}_1}^2-{\tilde{k}_2}^2}\nonumber\\\\
&&R^{(1)}(\textbf{x})\,=\,
-\frac{{m_R}^2}{f_R(0,\phi^{(0)})}\frac{r_g}{|\textbf{x}|}\frac{({\tilde{k}_1}^2-\eta^2)\,e^{-m_R\tilde{k}_1\,|\textbf{x}|}-({\tilde{k}_2}^2-\eta^2)\,e^{-m_R\tilde{k}_2\,|\textbf{x}|}}{{\tilde{k}_1}^2-{\tilde{k}_2}^2}\nonumber
\end{eqnarray}
where $r_g$ is the Schwarzschild radius. Furthermore,  we introduced the dimensionless quantities 

\begin{eqnarray}
2\,{\tilde{k}_{1,2}}^2\,=\,\frac{2\,{k_{1,2}}^2}{{m_R}^2}\,=\,1-\xi+\eta^2\pm\sqrt{(1-\xi+\eta^2)^2-4\eta^2}
\end{eqnarray}
and ${\displaystyle \eta\,=\,\frac{m_\phi}{m_R}}$, ${\displaystyle \xi\,=\,\frac{3{f_{R\phi}(0,\phi^{(0)})}^2}{2\,\omega(\phi^{(0)})\,f_R(0,\phi^{(0)})}}$. These two parameters  have to ensures two different conditions for the roots ${k_{1,2}}^2$ that have to be both real and positive \emph{i.e.} ${k_{1,2}}^2\,>\,0$. Such conditions can be reformulated as $\xi\,\leq\,(\eta-1)^2$ and this fact restrict the class of viable Lagrangians. In fact we have

\begin{eqnarray}
\xi\,=\,\frac{({k_1}^2-{m_R}^2)({k_2}^2-{m_R}^2)}{{m_R}^4}\,,\,\,\,\,\,\,\,\,\,\,
\eta^2\,=\,\frac{{k_1}^2{k_2}^2}{{m_R}^4}
\end{eqnarray}
where $\xi$ and $\eta$ are given in terms of ${k_{1,2}}^2$ and $m_R$ which are the parameters defining the form of Yukawa-like terms in the potentials. Specifically the conditions

\begin{eqnarray}
&&f_{RR}(0,\phi^{(0)})\,=\,-\frac{f_R(0,\phi^{(0)})}{3\,{m_R}^2}\,,\,\,\,\,\,\,\,\,\,\,f_{\phi\phi}(0,\phi^{(0)})\,=\,-2\,\omega(\phi^{(0)})\,\frac{{k_1}^2{k_2}^2}{{m_R}^2}\,,\nonumber\\\\
&&f_{R\phi}(0,\phi^{(0)})\,=\,\sqrt{\frac{2}{3}\omega({\phi^{(0)}})f_R(0,\phi^{(0)})\biggl[\frac{({k_1}^2-{m_R}^2)({k_2}^2-{m_R}^2)}{{m_R}^4}\biggr]}\nonumber
\end{eqnarray}
which, together with conditions (\ref{PPN-field-equation-general-theory-fR-O0}), give the form of possible Lagrangians. It is worth noticing that $f_R(0,\phi^{(0)})$ can be assumed equal to $1$ in standard units, while $\omega(\phi^{(0)})$ fixes the form of scalar field kinetic term that is equal to $1/2$ in the canonical case. For $\omega(\phi^{(0)})\,<\,0$ a ghost scalar field is possible.

\begin{figure}[htbp]
  \centering
  \includegraphics[scale=1]{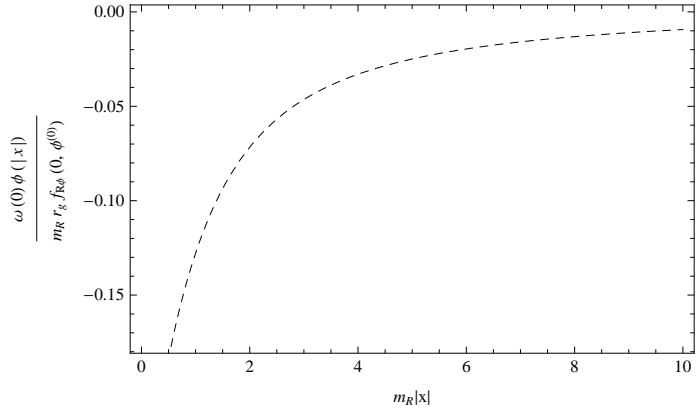}\\
  \caption{The spatial behavior of scalar field $\phi^{(1)}$ generated by a point-like source (\ref{coupledsyst_sol_point}) for $\eta\,=\,0.1$ and $\xi\,=\,-2$.}
  \label{plotscalar}
\end{figure}

\begin{figure}[htbp]
  \centering
  \includegraphics[scale=1]{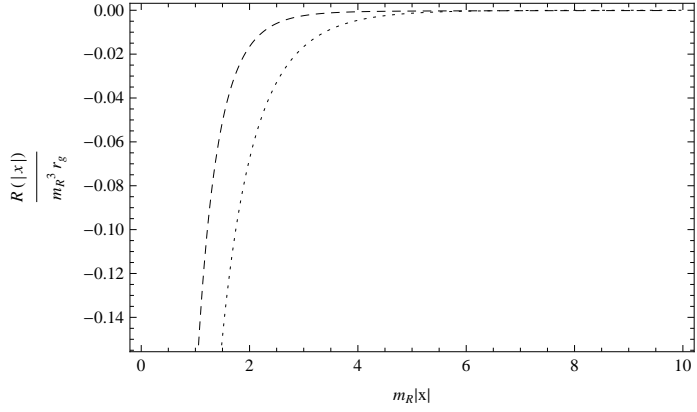}\\
  \caption{The spatial behavior of scalar field $R^{(1)}$ (dashed line) generated by a point-like source (\ref{coupledsyst_sol_point}) compared with respect to the same quantity (dotted line) in the $f(R)$-gravity. In both cases we set $\eta\,=\,0.1$ and $\xi\,=\,-2$.}
  \label{plotricciscalar}
\end{figure}

\begin{figure}[htbp]
  \centering
  \includegraphics[scale=1]{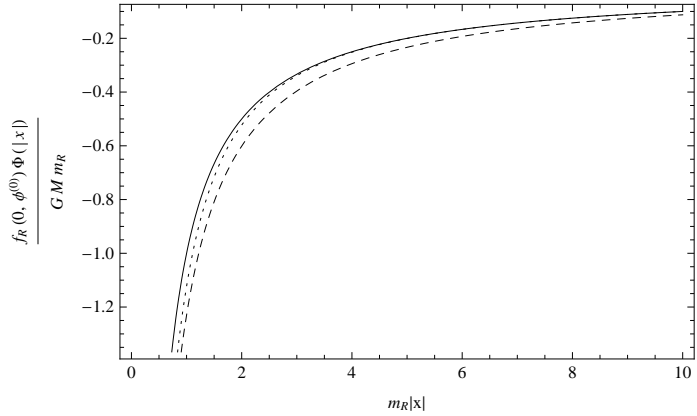}\\
  \caption{Comparison among the potentials $\Phi$ generated by a point-like source in three frameworks: the first one (\ref{new_sol_point}) induced by the action (\ref{HOGaction}) (dashed line), the second one (dotted line) induced by $f(R)$-gravity (\ref{new_sol_point_fR}) and the last one (solid line) is the Newtonian limit of GR. In the two alternative theories,  we set $\eta\,=\,0.1$ and $\xi\,=\,-2$.}
  \label{plotpotential}
\end{figure}

\begin{figure}[htbp]
  \centering
  \includegraphics[scale=1]{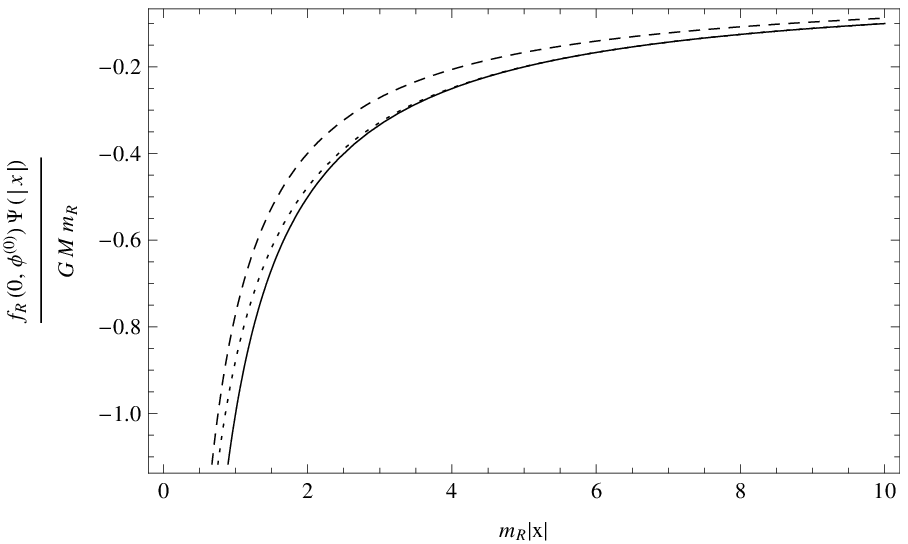}\\
  \caption{Comparison among the potentials $\Psi$ generated by a point-like source in three frameworks. The scheme is the same of Fig. \ref{plotpotential}.}
  \label{plotpotential_2}
\end{figure}

The potentials (\ref{new_sol}) become

\begin{eqnarray}\label{new_sol_point}
\Phi(\mathbf{x})\,&=&\,-\frac{GM}{f_R(0,\phi^{(0)})|\textbf{x}|}\biggl\{1+g(\xi,\eta)\,e^{-m_R\tilde{k}_1\,|\textbf{x}|}+[1/3-g(\xi,\eta)]\,e^{-m_R\tilde{k}_2\,|\textbf{x}|}\biggr\}\nonumber\\\\
\Psi(\mathbf{x})\,&=&\,-\frac{GM}{f_R(0,\phi^{(0)})|\textbf{x}|}\biggl\{1-g(\xi,\eta)\,e^{-m_R\tilde{k}_1\,|\textbf{x}|}-[1/3-g(\xi,\eta)]\,e^{-m_R\tilde{k}_2\,|\textbf{x}|}\biggr\}\nonumber
\end{eqnarray}
where ${\displaystyle g(\xi,\eta)\,=\,\frac{{\tilde{k}_1}^2(2\eta^2-2{\tilde{k}_1}^2-2\xi+3)-3\eta^2}{3{\tilde{k}_1}^2({\tilde{k}_1}^2-{\tilde{k}_2}^2)}}$. In Figs. (\ref{plotscalar}), (\ref{plotricciscalar}), (\ref{plotpotential}) and (\ref{plotpotential_2}) we show respectively the spatial behaviors of the scalar field, the Ricci scalar and the potentials $\Phi$, $\Psi$.


The solutions (\ref{coupledsyst_sol_point}), (\ref{new_sol_point}) are the generalization of solutions obtained in  $f(R)$-gravity and scalar tensor gravity. In fact, we can easily  obtain the outcomes that  in the case of minimally coupled scalar field \emph{i.e.} $f_{R\phi}\,=\,0\,\rightarrow\,\xi\,=\,0$, ${\tilde{k}_{1,2}}^2\,=\,1,\,\eta^2$, $g(\xi,\eta)\,=\,1/3$ we find the point-like solutions of $f(R)$-gravity \cite{PRD1, PRD1_2}

\begin{eqnarray}\label{new_sol_point_fR}
\Phi_{f(R)}(\mathbf{x})\,&=&\,-\frac{GM}{f_R(0)|\textbf{x}|}\biggl\{1+\frac{1}{3}\,e^{-m_R\,|\textbf{x}|}\biggr\}\nonumber\\\\
\Psi_{f(R)}(\mathbf{x})\,&=&\,-\frac{GM}{f_R(0)|\textbf{x}|}\biggl\{1-\frac{1}{3}\,e^{-m_R\,|\textbf{x}|}\biggr\}\nonumber
\end{eqnarray}
 while in the case of the Brans-Dicke theory \emph{i.e.} $f_{R\phi}=1,\,\,f_{R}=\phi,\,\,\omega(\phi)\,=\,-\omega_0/\phi,\,\,f_{RR}\,=\,0,\,\,f_{\phi\phi}\,=\,0$, from the field Eqs. (\ref{NL-fieldequationHOG}),  we find the classical solutions of Brans-Dicke gravity \cite{BD}
\begin{eqnarray}\label{sol_point_BD}
\Phi_{BD}(\mathbf{x})\,&=&\,-\frac{GM}{\phi^{(0)}|\textbf{x}|}\frac{2(2+\omega_0)}{2\,\omega_0+3}
\nonumber\\\\
\Psi_{BD}(\mathbf{x})\,&=&\,-\frac{GM}{\phi^{(0)}|\textbf{x}|}\frac{2(1+\omega_0)}{2\,\omega_0+3}
\nonumber
\end{eqnarray}
The Brans-Dicke behavior is also present in the solutions (\ref{new_sol_point}). In fact we recover the solutions (\ref{sol_point_BD}) when $m_R\,\rightarrow\,\infty,\,\,m_\phi\,=\,0,\,\,{\displaystyle \xi\,=\,-\frac{3}{2\,\omega_0}}$. Also solutions (\ref{new_sol_point_fR}) have been found in  vacuum, but here also the boundary conditions for the Yukawa term in the origin (where the mass is placed)  have been inserted. In  this case, the arbitrary time-function $\delta_1(t)$ in  Eq. (\ref{gravpot}) is fixed to the value ${\displaystyle  -\frac{2 m^2}{f_1}}$  and then we can define  $\Phi_{grav}\,=\,\Phi_{f(R)}$, while the expression for $\Psi_{grav}$ in the Eq. (\ref{mesol}) becomes

\begin{eqnarray}\label{pot_grav_2}
\Psi_{grav}\,=\,\frac{g^{(1)}_{rr}}{2}\,=\,-\frac{GM}{f_1r}\biggl\{1-\frac{1+mr}{3}\,e^{-mr}\biggr\}
\end{eqnarray}
It is possible to show that the potential $\Psi_{grav}$ is equal to the potential $\Psi_{f(R)}$ of Eqs. (\ref{new_sol_point_fR}) if we assume standard coordinates. The passage from the isotropic coordinates $(t,x^1,x^2,x^3)$ to the standard ones $(t,r,\theta,\phi)$ is given by the transformation $\biggl[1-2\,\Psi_{f(R)}(|\textbf{x}|)\biggr]|\textbf{x}|^2\,=\,r^2$ where $|\textbf{x}|^2\,=\,x_ix^i$, then, at first order with respect to the quantity $r_g/r$ (or $r_g/|\textbf{x}|$), the metrics 

\begin{eqnarray}\label{isotropic_metric}
 &&ds^2\,=\,\biggl[1-\frac{r_g}{f_R(0)|\textbf{x}|}\biggl(1+\frac{1}{3}\,e^{-m_R|\textbf{x}|}\biggr)\biggr]dt^2-\biggl[1+\frac{r_g}{f_R(0)|\textbf{x}|}
  \biggl(1-\frac{1}{3}\,e^{-m_R|\textbf{x}|}\biggr)\biggr]\delta_{ij}dx^idx^j\nonumber\\\\
&&ds^2\,=\,\biggl[1-\frac{r_g}{f_1\,r}\biggl(1+\frac{1}{3}\,e^{-mr}\biggr)\biggr]dt^2-\biggl[1+\frac{r_g}{f_1\,r}
  \biggl(1-\frac{1+mr}{3}\,e^{-mr}\biggr)\biggr]dr^2-r^2d\Omega\nonumber
\end{eqnarray}
coincide and, obviously, for $f(R,\phi)\,\rightarrow\,f(R)$, we have $m_R\,=\,m$ and $f_R(0)\,=\,f_1$.

It is interesting to note that, in the case of minimally coupled scalar field ($f_{R\phi}\,=\,0$),  the Newtonian level of the field Eqs. (\ref{fieldequationHOG}) for the metric tensor is unaffected by the presence of the scalar field $\phi$. Moreover $\phi$ is not linked to the energy-momentum tensor \emph{via} the Ricci scalar and must satisfy only the boundary conditions at infinity, while the amplitude of scalar field is generic and depending only on  time. In fact, the solution of the  first of  Eqs. (\ref{coupledsyst}) is

\begin{eqnarray}
\phi^{(1)}(t,\textbf{x})\,=\,\frac{K(t)}{2\,\omega(\phi^{(0)})}\frac{e^{-m_\phi|\textbf{x}|}}{|\textbf{x}|}\,,
\end{eqnarray}
where $K(t)$ is a generic function  depending on time. The evolution of $K(t)$ is fixed by the post-Newtonian level of field equations. By  considering $f_{R\phi}\,\neq\,0$,  we find a further contribution in the energy-momentum tensor. Another interesting case is the generalization of Brans-Dicke theory. In fact,  we can consider a scalar-tensor theory, but the geometric sector is given only by the Ricci scalar. Without losing  generality,  we can set the interaction term in the action (\ref{HOGaction}) as $\phi\,R$ and not as $f(\phi)\,R$ because by introducing a new scalar field we obtain formally the same equations \cite{TS_analogy}. By setting  $f_{R\phi}\,=\,1,\,\,f_{RR}\,=\,0,\,\,f_R\,=\,\phi$ the field Eqs.  (\ref{NL-fieldequationHOG}) become\footnote{Also in this case we can find the solution of $\Psi$ as $\Psi(\textbf{x})\,=\,\frac{1}{8\pi}\int
d^3\textbf{x}'\frac{R^{(1)}(\textbf{x}')}{|\textbf{x}-
\textbf{x}'|}-\frac{\phi^{(1)}(\textbf{x})}{\phi^{(0)}}$.}

\begin{eqnarray}
\label{NL-fieldequationST}
&&\triangle\Phi\,=\,\frac{\mathcal{X}\,\rho}{\phi^{(0)}}+\frac{\triangle\phi^{(1)}}{\phi^{(0)}}+\frac{R^{(1)}}{2}\nonumber\\\nonumber\\
&&\Psi\,=\,\Phi+\frac{\phi^{(1)}}{\phi^{(0)}}\nonumber\\\\
&&\biggl[\triangle-{m_\phi}^2\biggr]\phi^{(1)}\,=\,-\frac{R^{(1)}}{2\,\omega(\phi^{(0)})}\nonumber\\
&&R^{(1)}\,=\,-\frac{\mathcal{X}\,\rho}{\phi^{(0)}}-\frac{3\,\triangle\phi^{(1)}}{\phi^{(0)}}\nonumber
\end{eqnarray}
and their solutions are

\begin{eqnarray}
\label{NL-solution_ST}
&&\phi^{(1)}(\textbf{x})\,=\,-\frac{1}{2\,\omega(\phi^{(0)})\,\phi^{(0)}-3}\frac{r_g}{|\textbf{x}|}\,e^{-\sqrt{\frac{2\,\omega(\phi^{(0)})\,\phi^{(0)}}{2\,\omega(\phi^{(0)})\,\phi^{(0)}-3}}\,m_\phi |\textbf{x}|}\nonumber\\\nonumber\\
&&R^{(1)}(\textbf{x})\,=\,-\frac{4\pi\,r_g}{\phi^{(0)}}\,\delta(\textbf{x})+\frac{6\,\omega(\phi^{(0)})\,{m_\phi}^2}{[2\,\omega(\phi^{(0)})\,\phi^{(0)}-3]^2}\frac{r_g}{|\textbf{x}|}\,e^{-\sqrt{\frac{2\,\omega(\phi^{(0)})\,\phi^{(0)}}{2\,\omega(\phi^{(0)})\,\phi^{(0)}-3}}\,m_\phi |\textbf{x}|}\nonumber\\\\
&&\Phi_{ST}(\textbf{x})\,=\,-\frac{GM}{\phi^{(0)}|\textbf{x}|}\biggl\{1-\frac{e^{-\sqrt{\frac{2\,\omega(\phi^{(0)})\,\phi^{(0)}}{2\,\omega(\phi^{(0)})\,\phi^{(0)}-3}}\,m_\phi |\textbf{x}|}}{2\,\omega(\phi^{(0)})\,\phi^{(0)}-3}\biggr\}\nonumber\\\nonumber\\
&&\Psi_{ST}(\textbf{x})\,=\,-\frac{GM}{\phi^{(0)}|\textbf{x}|}\biggl\{1+\frac{e^{-\sqrt{\frac{2\,\omega(\phi^{(0)})\,\phi^{(0)}}{2\,\omega(\phi^{(0)})\,\phi^{(0)}-3}}\,m_\phi |\textbf{x}|}}{2\,\omega(\phi^{(0)})\,\phi^{(0)}-3}\biggr\}\nonumber
\end{eqnarray}
which, in the case of massless scalar field, become the typical solution of Brans-Dicke theory.

In the  cases that we have  shown,  the Newtonian contribution ($|\textbf{x}|^{-1}$) to the potential is ever present. We can find a difference in the definition of gravitational constant $G$, since in these theories we have a multiplying factor $f_R(0,\phi^{(0)})^{-1}$, while the additional terms are depending on the form of the Lagrangian. It is important to stress that, in all cases that  we have considered,  the limit and results of   GR  are fully recovered. 

If we have a generic matter source distribution $\rho(\textbf{x})$, it is sufficient to use the superposition principle by starting from  point-like solutions. Then we substitute to the solutions (\ref{new_sol_point}) the integral expression: $\Phi\,\rightarrow\,\int\Phi$. This approach is correct only in the Newtonian limit since  such a limit correspond also to the linearized version of the theory. 


\section{Rotation curves of galaxies}
\label{cinque}

At astrophysical level,  the probe for the validity of alternative theories of gravity is the correct reproduction of   rotation curves of spiral galaxies \cite{annalen}. As discussed above,  the  foundation of the dark matter issue lies on this observational evidence. 
In order to face such a problem, one has to discuss the motion of a  body embedded in a gravitational field. Let us take into account the  geodesic equation

\begin{eqnarray}\label{geodesic}
\frac{d^2\,x^\mu}{ds^2}+\Gamma^\mu_{\alpha\beta}\frac{dx^\alpha}{ds}\frac{dx^\beta}{ds}\,=\,0
\end{eqnarray}
where $ds\,=\,\sqrt{g_{\alpha\beta}dx^\alpha dx^\beta}$ is the relativistic distance. In the Newtonian limit,   from Eq.(\ref{geodesic}), we obtain    the equation of motion equation

\begin{eqnarray}
\frac{d^2\,\mathbf{x}}{dt^2}\,=\,-\nabla\Phi(\mathbf{x})\,.
\end{eqnarray}
In our case,  the gravitational potentials are given by (\ref{new_sol_point}). The study of motion is very simple if we consider a particular symmetry of mass distribution $\rho$, otherwise the analytical solutions are not available. Our aim is to evaluate the corrections to the classical motion in the easiest situation: the circular motion. In this case we do not consider the radial and vertical motions. The condition of stationary motion on the circular orbit is

\begin{eqnarray}\label{stazionary_motion}
v_c(|\mathbf{x}|)\,=\,\sqrt{|\mathbf{x}|\frac{\partial\Phi(\mathbf{x})}{\partial|\mathbf{x}|}}
\end{eqnarray}
where $v_c$ is the velocity. Generally the correction terms do not satisfy the Gauss theorem \cite{Stabile_Capozziello} and this aspect implies that a sphere cannot be simply reduced to a point. In fact the gravitational potential generated by a sphere (also with constant density) is depending also on the Fourier transform of the sphere \cite{Stabile_Capozziello}. Only in the limit case, where the radius of the sphere is small with respect to the distance (point-like source), we obtain the simple expression (\ref{new_sol_point}).

\begin{figure}[htbp]
  \centering
  \includegraphics[scale=1]{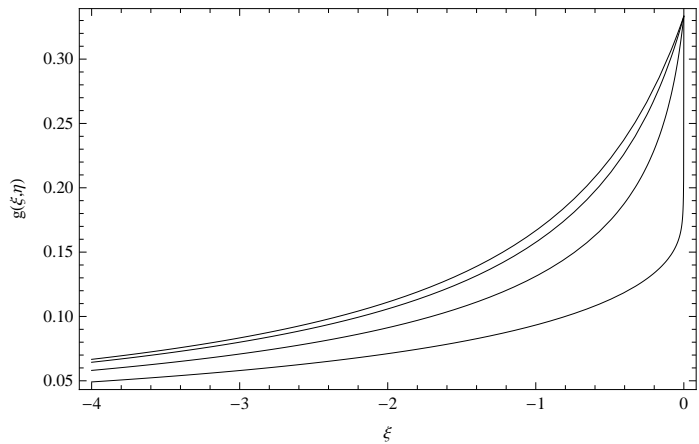}\\
  \caption{Plot of coefficient $g(\xi,\eta)$ with respect to quantity $\xi$ for $0\,\leq\eta\,\leq\,0.99$ with step $0.33$.}
  \label{plotcoefficient_1}
\end{figure}

A further remark on Eqs. (\ref{new_sol_point}) is needed. The structure of solutions is mathematically similar to the one of fourth-order gravity generated by $f(R,R_{\alpha\beta}R^{\alpha\beta})$. However  there is a fundamental difference with the present case: the two Yukawa corrections have different algebraic sign. In particular, the Yukawa correction induced by a generic function of Ricci scalar  implies a stronger attractive  gravitational force, while the second one, induced by squared Ricci tensor, implies a repulsive force \cite{PRD1, stabile_scelza}. In the present paper, the Yukawa corrections are induced by a generic function of Ricci scalar and  a non-minimally coupled scalar field. Both corrections have a positive coefficient. In fact in Fig. (\ref{plotcoefficient_1}), we show the coefficient $g(\xi,\eta)$ with respect to $\xi$ for a given values of $\eta$. The function $g(\xi,\eta)$ assumes the maximum value ($\,=\,1/3$) when $\xi\,=\,0$ (we have a pure $f(R)$-gravity and the scalar field does not contribute to the gravitational potential in this case), otherwise we have two Yukawa corrections with positive coefficients. The scalar field gives rise to  a more attractive force than in  $f(R)$-gravity. The interesting range of values of $\eta$ is between $0$ and $1$. In the case $\eta\,>\,1\,\rightarrow\,m_\phi\,>\,m_R$, the correction induced by scalar field is suppressed with respect to the other one.

\begin{figure}[htbp]
  \centering
  \includegraphics[scale=1]{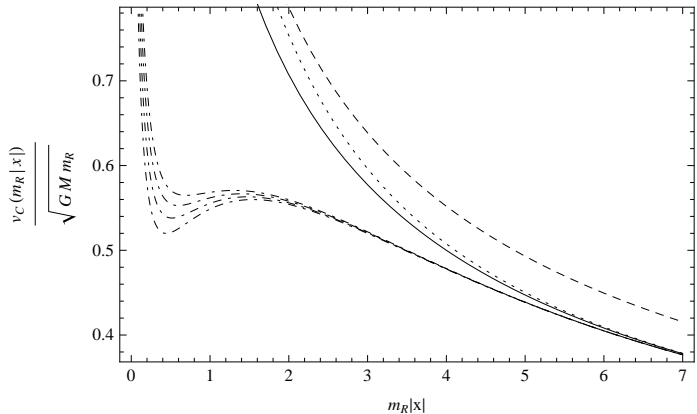}\\
  \caption{Comparison (in the vacuum case)  of  the spatial behaviors of rotation curves in the gravitational field generated by a point-like source. The dotted and dashed lines are the Sanders model for $-0.95\,<\,\alpha\,<\,-0.92$, the solid line is the GR curve, the dotted line is the $f(R)$-gravity and the dashed line is the fourth order gravity non-minimally coupled with a scalar field. In the last case,  we set $\xi\,=\,-2$, $\eta\,=\,0.1$.}
  \label{plotcircul}
\end{figure}

From this analysis,  the choice of $f(R,\phi)$-gravity  is better than  $f(R,R_{\alpha\beta}R^{\alpha\beta})$-gravity, but we have a problem in the limit for $|\textbf{x}|\,\rightarrow\,\infty$: the interaction is, of course, scale-depending (the scalar fields are massive) and in the vacuum the corrections turn off. For this reason,  at large distances, we recover only the classical Newtonian contribution. Therefore the presence of scalar fields makes the  profile smooth. This behavior is very clear in the study of rotation curves (\ref{stazionary_motion}). Let us  assume  a phenomenological   point-like gravitational potential as supposed by Sanders \cite{Sanders90,sanders}

\begin{eqnarray}\label{san_pot}
\Phi_{SP}(\textbf{x})\,=\,-\frac{GM}{|\textbf{x}|}(1+\alpha\,e^{-m|\textbf{x}|})
\end{eqnarray}
where $\alpha$ and $m$ are  free parameters that, following Sanders \cite{sanders}, can be assumed to be $\alpha\,\simeq\,-0.92$ and $r_0\,=\,1/m\,\simeq\,40\, \text{Kpc}$ to fit the galactic rotation curves. This potential has been introduced to explain the rotation curves of spiral galaxies \cite{sanders,cardoneyoukawa}, however the theoretical framework generating it is purely phenomenological.  Recently by using the same potential it has been possible to fit elliptical galaxies \cite{cap_de_na}. In both cases by setting a negative value to $\alpha$ an almost  constant profile of rotation curve is recovered. Such a  rotation curve is obviously possible but there are two problems: the first one consists that no $f(R, \phi)$-gravity, by imposing all boundary conditions at origin and at infinity, gives  that negative value of $\alpha$. The second one is linked to the value of gravitational constant $G$. In fact in presence of Yukawa-like correction with negative coefficient,  we find a lower rotation curve and only by resetting $G$ (or the point-like mass) we can fit the experimental data. In Fig. (\ref{plotcircul}) we compare the profiles derived in the Newtonian limit of  GR, $f(R)$- and $f(R,\phi)$-gravity and the potential (\ref{san_pot}). It is extremely interesting to note that the presence of scalar the field $\phi$, in the case $m_\phi\,\sim\,m_R$, guarantees a rotation curve higher than the other ones but also in this case we find the  asymptotic flatness are derived from observations. 
Only if we consider a massive scalar tensor theory non-minimally coupled,  we get a potential with negative coefficient in Eq.(\ref{NL-solution_ST}). In fact by setting the gravitational constant as ${\displaystyle G_0\,=\,\frac{2\,\omega(\phi^{(0)})\,\phi^{(0)}-4}{2\,\omega(\phi^{(0)})\,\phi^{(0)}-3}\frac{G_\infty}{\phi^{(0)}}}$ where $G_\infty$ is the gravitational constant as measured at infinity and by imposing $\alpha^{-1}\,=\,3-2\,\omega(\phi^{(0)})\,\phi^{(0)}$, the potential $\Phi_{ST}$ in the (\ref{NL-solution_ST}) becomes

\begin{eqnarray}
\label{ST_pot}
\Phi_{ST}(\textbf{x})\,=\,-\frac{G_\infty M}{|\textbf{x}|}\biggl\{1+\alpha\,e^{-\sqrt{1-3\alpha}\,m_\phi |\textbf{x}|}\biggr\}
\end{eqnarray}
and then the Sanders potential  (\ref{san_pot}) is fully recovered.

\section{Discussion and Conclusions}
\label{sei}
The dark matter issue, together with dark energy, can be considered  the major problem of modern astrophysics and cosmology. Beside the huge amount of observations confirming its effects,  practically at all astrophysical scales, no final answer exists, at fundamental level, definitively confirming one (or more than one) candidate supposed to explain the phenomenology. Furthermore, GR has been firmly tested only up to Solar System scale and then its features have been inferred at larger scales. In this situation, dark matter and dark energy could be nothing else but the manifestation that GR does not work at IR scales.

 A similar disturbing situation is found at UV scales where no Quantum Gravity theory is up to now definitely available. Alternative gravities (in particular ETGs) could represent a way out to this puzzle being  effective theories of gravity  representing a reliable picture of quantum fields in high curvature regimes \cite{libro} and an approach to overcome the dark side problem at larger scales. 

In this paper, we have discussed the weak field limit (in particular the Newtonian limit) of some classes of ETGs in view to explain the almost flat rotation curves of spiral galaxies. In particular, we have shown that ETGs, in general,  present Yukawa-like corrections in the gravitational potential. In particular, we have analyzed the case of $f(R)$ and $f(R,\phi)$. The latter are known to be analogue to $f(R,\Box R)$. 

After a discussion of the mathematical features of the emerging corrections, we have confronted the results with the phenomenological Sanders potential, assumed as a possible dynamical explanation of flat rotation curves. The suitable value of phenomenological parameters can be exactly reproduced in the framework of 
$f(R,\phi)$-gravity since the concurring Yukawa corrections allow to recover attractive and repulsive components of potential. In this case, no dark matter is required to fit dynamics like in the case discussed in \cite{cardoneyoukawa} where only a Yukawa-like correction was not sufficient to reproduce realistic rotation curves.

\end{document}